\documentclass[preprint,floats,aps,epsfig,nofootinbib,amssymb]{revtex4}
\usepackage{graphicx}
\usepackage{dcolumn}
\usepackage{bm}
\usepackage{mathrsfs}
\usepackage{slashed}
\usepackage{graphicx,color}
\usepackage{epsfig}
\usepackage{subfigure}
\usepackage{CJK}
\usepackage{xcolor}
\usepackage{amsmath}
\usepackage{amssymb}
\usepackage{caption2}
\usepackage{float}
\usepackage{amsmath}
\usepackage{amsmath}
\usepackage{hyperref}
\begin{document}

\title{Study on direct pion emission in decay $D^{*+} \to D^+ \pi$ }
\author{Xing-Dao Guo$^1$}
\email{guoxingdao@mail.nankai.edu.cn}
\author{Xuewen Liu$^1$}
\email{liuxuewen@mail.nankai.edu.cn}
\author{Hong-Wei Ke$^2$}
\email{khw020056@tju.edu.cn}
\author{Xue-Qian Li$1$}
\email{lixq@nankai.edu.cn}
\affiliation{1. School of Physics, Nankai University, Tianjin, 300071, P.R.~China\\
2. Department of Physics, Tianjin University, Tianjin, 300072, P.R. China}
%\affiliation{1. School of Physics, Nankai University, Tianjin 300071, P. R. China \\2. \\}

\begin{abstract}
The QCD multipole expansion (QCDME) is based on the quantum field theory, so should be more reliable. However,
on another aspect, it refers to the non-perturbative QCD , so that has a certain application range. Even though
it successfully explains the data of transition among members of the $\Upsilon$ ($\psi$) family, as Eichten indicates,
beyond the production threshold of mediate states it fails to meet data by several orders. In this work, by studying a
simple decay mode $D^*\to D+\pi^0$, where a pion may be emitted before $D^*$ transiting into $D$, we analyze the contribution
of QCD multipole expansion. Whereas as the $D\pi$ portal is open, the dominant contribution is an OZI allowed process
where a light quark-pair is excited out from vacuum and its contribution can be evaluated by the $^3P_0$ model. Since
the direct pion emission is a process which is OZI suppressed and violates the isospin conservation, its contribution must be much smaller
than the dominant one. By a careful calculation, we may quantitatively estimate how small the QCDME contribution should be
and offer a quantitative interpretation for Eichten's statement.

\end{abstract}

\maketitle

\section{Introduction}

The QCD Multipole Expansion (QCDME) has been widely used to calculate transition rates among heavy quarkonia by emitting pions,
\cite{Yan:1980uh,Kuang:1981se}. Since this theory refers to non-perturbative QCD,
it has a limited application range, beyond this range the theory is no longer applicable. When the masses of the charmonia
(bottononia) are sufficiently large beyond the production thresholds of  $D^{(*)}\bar D^{(*)}$($(B^{(*)}\bar B^{(*)})$) which may become
on-shell intermediate states, as Eichtein et al. indicate, the decay widths evaluated in terms of the QCDME are smaller than the
data by three orders\cite{eichten2013}. In other words, the dominant modes of, say, $\Upsilon(nS)\to \Upsilon(mS)+\pi^+\pi^-$ or
$\Upsilon(nS)\to \Upsilon(mS)+\pi^0$ can be realized via $\Upsilon(nS)\to B^{(*)}\bar B^{(*)}\to \Upsilon(mS)+\pi^+\pi^-$, which is usually
referred as the final state interaction or re-scattering process. Even though the re-scattering process dominates the transition, the direct
pion emission is still contributing and is evaluated in terms of the QCDME. It is interesting to theoretically estimate how small the
contribution of the direct pion emission could be in comparison with the dominant one.

To serve the purpose, we adopt a simple decay mode to do the job, i.e. calculate the contribution of a direct $\pi^0$ emission to the decay rate of
$D^*\to D+\pi^0$.

For $D^{*+} \to D^+ \pi^0$, the direct $\pi^0$ emission is an OZI suppressed and moreover causes an isospin violation. The double suppression
determines that the contribution from the direct pion emission must be small. In fact, unless
other mechanisms are forbidden by some reasons, such as constraints of available phase space or other symmetries, the direct pion emission
cannot make substantial contribution to the decay rates as Eichten et al. suggest. To quantitatively confirm Eichten's statement,
we use both the $^3P_0$ model and
QCDME to calculate their contribution to the decay rate separately.
Our numerical results show that the
effective coupling constant $g_{D^*D\pi}$
determined by QCDME is 60$\sim$ 70 times smaller than that obtained from quark pair creation (QPC).

After the introduction, in section \ref{s2}, we evaluate the
contributions to the decay rate of the  $D^{*+} \to D^+ \pi^0$
from both the quark pair creation (QPC) described by the  $^3P_0$
model and the direct pion emission described by the QCDME
respectively in subsections \ref{ss21} and \ref{ss22}. The numerical results are
presented following the formulations in the section and
comparisons with the corresponding experimental data are made.
% we briefly review the QCDME and then give the equation of the $\pi$ radiation through QCDME.
In the final section we will discuss the framework in some details and then draw our conclusion.

\section{$D^{*+} \to D^+ \pi^0$ decays\label{s2}}
\subsection{The quark pair creation model and its application to $\pi^0$ radiation\label{ss21}}

\begin{figure}[H]
\centering
\begin{minipage}[!htbp]{0.5\textwidth}
\centering
\includegraphics[width=0.98\textwidth]{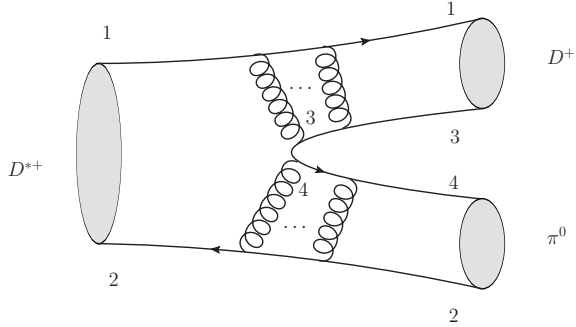}
\caption{The quark-pair creation from vacuum serves as the decay mechanism for  $D^{*+} \to D^+ \pi$. }
\label{3p0}
\end{minipage}
\end{figure}

In the framework of the QPC model\cite{Micu,yaouanc,yaouanc-1,yaouanc-book,Beveren,BSG,sb,qpc-1,qpc-2,qpc-90,
ackleh,Zou,liu,Close:2005se,lujie,xiangliu-2860,xiangliu-heavy,Li:2008mz}, the  decay  $D^{*+} \to D^+ \pi^0$
occurs via a quark-antiquark pair creation from the vacuum. It is an Okubo-Zweig-Iizuka (OZI) allowed process. The decay mechanism
is displayed in Fig.\ref{3p0} graphically. The picture is that many soft gluons are emitted from the quark and anti-quark legs which then
annihilate into a quark-antiquark pair.
Equivalently, the physics scenario can be described as that a quark pair is excited out from vacuum.
The $^3P_0$ model has been widely applied to calculate such hadronic strong decays.

%This model is applicable to Okubo-Zweig-Iizuka (OZI) allowed strong decays of
%a hadron into two other hadrons, which are expected to be the dominant decay modes
%of a meson if they are allowed.

%The decay $D^{*+} \to D^+ \pi^0$ occurs by emitting a $\pi^0$.

For readers' convenience, we collect the relevant formulations about the calculation in terms of the $^3P_0$ model in the appendix.
%The created quark pair is of the quantum number of the vacuum, $0^{++}$ \cite{Micu,yaouanc}.
%In the non-relativistic limit,
The transition operator for the quark pair creation reads
\begin{eqnarray}
T&=& - 3 \gamma \sum_m\: \langle 1\;m;1\;-m|0\;0 \rangle\,
\int\!{\rm d}{\textbf{k}}_3\; {\rm d}{\textbf{k}}_4
\delta^3({\textbf{k}}_3+{\textbf{k}}_4) {\cal
Y}_{1m}\left(\frac{{\textbf{k}}_3-{\textbf{k}_4}}{2}\right)\;
\nonumber\\&&\times\chi^{3 4}_{1, -\!m}\; \varphi^{3 4}_0\;\,
\omega^{3 4}_0\; d^\dagger_{3i}({\textbf{k}}_3)\;
b^\dagger_{4j}({\textbf{k}}_4)\,, \label{tmatrix}
\end{eqnarray}
and the hadronic matrix element is determined as
\begin{eqnarray}
\langle D^{+}\pi^0|S|D^{*+}\rangle=I-i2\pi\delta(E_{\text{final}}-E_{\text{initial}})\langle D^{+}\pi^0|T|D^{*+}\rangle.\label{smatrix}
\end{eqnarray}
In Eq.~\ref{tmatrix}, $i$ and $j$ are the SU(3)-color indices of the created quark and anti-quark. $\varphi^{34}_{0}=(u\bar u +d\bar d +s \bar
s)/\sqrt 3$ and $\omega_{0}^{34}=\delta_{ij}$ are for flavor and color singlets, respectively.
$\chi_{{1,-m}}^{34}$ is the spin wave function. $\mathcal{Y}_{\ell m}(\mathbf{k})\equiv
|\mathbf{k}|^{\ell}Y_{\ell m}(\theta_{k},\phi_{k})$ is the $\ell$th solid harmonic polynomial.
$\gamma$ is a dimensionless constant which denotes the strength of quark pair creation from
vacuum and is fixed by fitting data. Following Ref.\cite{Blundell:1995ev}, we take $\gamma=13.4$ in this work. For Eq.~\ref{smatrix}, the
explicit expressions for the wave function of a meson and the  hadronic matrix elements are presented in the appendix.

The helicity amplitude $\mathcal{M}^{M_{J_{D^{*+}}}M_{J_{D^{+}}}M_{J_{\pi^0}}}$ of this process can be
extracted from the relation $\langle D^{+}\pi^0|S|D^{*+}\rangle=\delta^3(\mathbf{K}_{D^{+}}+\mathbf{K}_{\pi^0}-\mathbf{K}_{D^{*+}})
\mathcal{M}^{M_{J_{D^{*+}}}M_{J_{D^{+}}}M_{J_{\pi^0}}}$. Then, the decay width corresponding to the process is written
in terms of the helicity amplitude as
\begin{eqnarray*}
\Gamma=\pi^2\frac{|\mathbf{K}|}{M_{D^{*+}}^2}\frac{1}{2J_{D^{*+}}+1}
\sum_{\renewcommand{\arraystretch}{.38}\begin{array}[t]{l}
\scriptstyle M_{J_{M_{D^{*+}}}},M_{J_{M_{D^{+}}}},
\scriptstyle  M_{J_{M_{\pi^0}}}
\end{array}}
\Big|\mathcal{M}^{M_{J_{D^{*+}}}M_{J_{D^{+}}}M_{J_{\pi^0}}}\Big|^2\,,
\end{eqnarray*}
where we take $\textbf{K}_{D^{+}}=-\textbf{K}_{\pi^0}=\textbf{K}$ in the center of the mass frame of  ${D^{*+}}$.
%, i.e. $\textbf{K}_{D^{*+}}=0$.

Numerically, we take a typical $R$ value for $D$ meson from Ref.~\cite{Close:2005se} as 2.3 GeV$^{-1}$ and  $R=2.1$ GeV$^{-1}$ for $\pi^0$ from
Ref.~\cite{Blundell:1995ev}. With these parameter setup, the decay width of $D^{*+} \to D^+ \pi^0$ can be easily obtained as $21.9$ keV.
Experimentally, the decay width of this mode is $29.5^{+7.3}_{-7.2}$ keV\cite{Agashe:2014kda}. The consistency of the numerical results evaluated
in terms  of the $^3P_0$ model with data indicates that the theoretical framework is well established and applicable to describe such processes.

%\subsection{Direct pion emission and the QCDME calculation}

On the other hand, based on the heavy quark effective theory(HQET), one can extract the effective
coupling constant of $D^*D\pi$ from the afore calculated decay width which might offer significant information for
application of an effective theory.  Following Ref.\cite{Casalbuoni:1996pg}, the related effective
Lagrangian can be written as
\begin{equation}
\begin{array}{rl}
\mathcal{L}=-\frac{2g_{D^*D\pi}}{f_\pi}D^*_\mu\partial^\mu\frac{\phi_\pi}{\sqrt 2}D^\dag+h.c.
\end{array}
\end{equation}
Then we can get the decay width as
\begin{equation}
\begin{array}{rl}
\Gamma(D^{*+} \to D^+ \pi^0)=\frac{1}{2m_{D^*}}\frac{4\pi|\textbf{k}|}{(2\pi)^2 4m_{D^*}}
\frac{|\mathcal{T}|^2}{3},
\end{array}
\label{decay}
\end{equation}
with $|\textbf{k}|=\frac{1}{{2m_{D^*}}}[(m^2_{D^*}-(m_\pi+m_{D})^2)(m^2_{D^*}-(m_\pi-m_{D})^2)]^{1/2}$. The transition amplitude of
$D^{*+} \to D^+ \pi^0$ is\cite{Casalbuoni:1996pg}
\begin{equation}
\begin{array}{rl}
\mathcal{T}(D^{*+} \to D^+ \pi^0)&=g_{D^*D\pi} \frac{1}{\sqrt 2}\frac{2m_D}{f_\pi}\textbf{k}\cdot\boldsymbol{\epsilon},
\end{array}
\end{equation}
here $\boldsymbol{\epsilon}$ is the polarization vector of $D^*$. From equation(\ref{decay}) we obtain $g_{D^*D\pi}=0.51$.

%%%%%%%%%%%%%%%%%%%%%%%%%%%%%%%%%%%%%%%%%%%%%%%%%%

\subsection{The QCDME and evaluating contribution of direct $\pi^0$ emission to the decay width\label{ss22}}

It is noted that the pion can be directly emitted before $D^*$ transits into $D$, thus the amplitude, in principle, should be added
to the process depicted in above subsection and interferes with it. Just by the qualitative analysis, the one-pion emission is an OZI suppressed
process and moreover, it violates isospin conservation, therefore must be much small compared to the vacuum creation.

%On the other aspect,
%in the process of $\Upsilon(nS)\to\Upsilon(mS)+|pi^0\; (n>m)$, the direct one-pion emission is the leading term which is accurately evaluated in terms
%of QCDME. However for higher excited states  $\Upsilon(nS)$ with larger $n$, the $B^{(*)}\bar B^{(*)}$ channels
%are open, the evaluated value by QCDME is smaller than data by several orders\cite{Eichten}. That threshold separates the two regions which are
%described by completely different mechanisms. It means that crossing the ridge, physics would be different.
It is obviously
interesting to investigate such an effect in other processes, i.e. as long as the one-pion emission is not a leading term, how small it would be
compared with the leading ones.   Below, we will quantitatively investigate the direct one-pion emission in $D^*\to D+\pi^0$.

\begin{figure}[H]
\centering
\begin{minipage}[!htbp]{1\textwidth}
\centering
\includegraphics[width=0.98\textwidth]{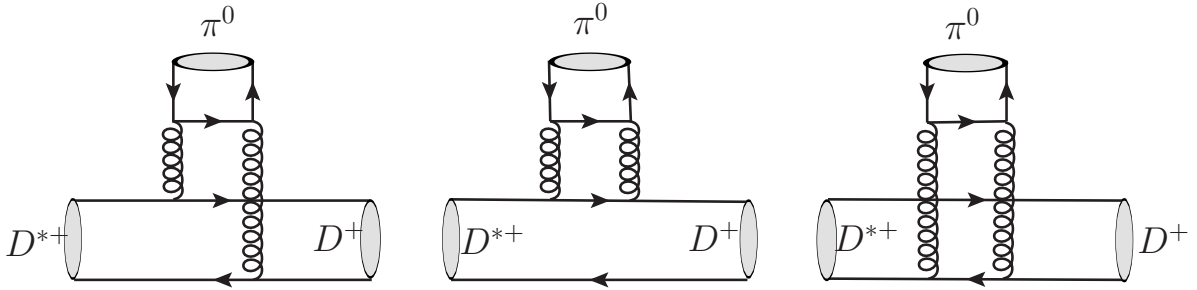}
\caption{The QCDME diagrams responsible for pion emission in the process $D^{*+} \to D^+ \pi^0$.}
\label{qcdme}
\end{minipage}
\end{figure}

The corresponding diagrams for the direct pion emission in $D^{*+} \to D^+ \pi^0$ for which the QCDME is responsible are shown in Fig.\ref{qcdme}.

\begin{figure}[H]
\centering
\begin{minipage}[!htbp]{1\textwidth}
\centering
\includegraphics[width=0.98\textwidth]{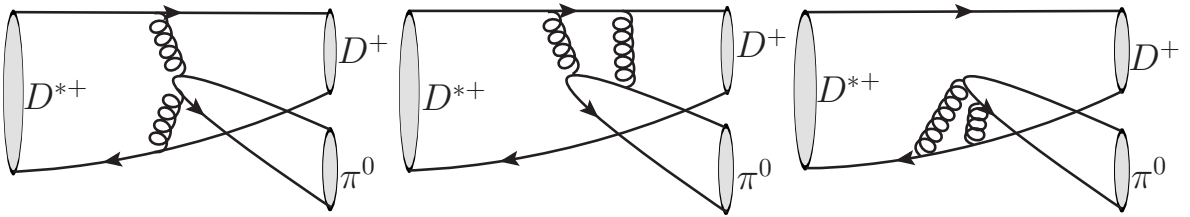}
\caption{Distortion of the $^3P_0$ decay mechanism for $D^{*+} \to D^+ \pi$ into an OZI suppressed process.}
\label{me3}
\end{minipage}
\end{figure}

The readers should note that Fig.\ref{qcdme} is just obtained by distorting Fig.\ref{3p0}, as shown in Fig.\ref{me3}.

Here we only
draw two gluon field lines, but as well understood, in the scenario of QCDME the lines correspond to a field of $E_n$ mode or $M_n$ one which are
by no means free gluons and the line is also not corresponding to a single-gluon propagator. Thus the line indeed
denote a collection of many soft gluons just as shown in Fig.\ref{3p0}.

Now let us calculate the rate contributed by the processes shown in Fig.\ref{qcdme} in the framework of QCDME.

%The multipole expansion in quantum chromodynamics has been widely used in studying
%the soft $\pi$/$\eta$ radiation of the charmonium and
%bottononium\cite{Yan:1980uh,Kuang:1981se,Kuang:1988bz,Kuang:1990kd,Kuang:2002hz,Kuang:2006me}. It expands gluon field in terms of $ka$, where $a$ is the
%size of heavy quarkonium and $k$ is the momentum of the radiated gluon.

The process of directly emitting a soft $\pi^0$ from $D^*$ in decay $D^*\to D+\pi^0$
is dominated by an E1-M2 coalesce transition. This is an OZI suppressed process and violates isospin conservation. The transition amplitude is\cite{Kuang:2006me}
\begin{equation}
\begin{array}{rl}
\mathcal{M}_{E1M2}=i\frac{g_E g_M}{12m}\sum_{NL}
(\frac{\langle\Phi_F|x_i|NL\rangle \langle NL|S_j x_k|\Phi_I\rangle}{M_I-E_{NL}}+
\frac{\langle\Phi_F|S_j x_k|NL\rangle \langle NL|x_i|\Phi_I\rangle}{M_I-E_{NL}})\langle\pi|E^a_i \partial_k B^a_j |0\rangle,
\end{array}
\end{equation}
where $S$ operator acting on the total spin of the heavy-quark and light-anti-quark system, $N$ and $L$ are the principal quantum number and the
orbital angular momentum of the intermediate hybrid state, $M_I$ and $E_{NL}$ are the mass of the initial meson $D^*$ and the energy eigenvalues
of the hybrid state, $m$ is the energy scale of the M2 transition and we set it to be $m_c$ and $\frac{m_c}{2}$ in our numerical computations.
The amplitude reduces into\cite{Kuang:1988bz,Kuang:1990kd}%$\mathbf{\tau},\boldsymbol{\tau},\pmb{\tau}$
\begin{equation}
\begin{array}{rl}
\mathcal{M}_{E1M2}=i\frac{g_E g_M}{18m}\sum_{NL}
\frac{\int R_F(r) r R_{NL}^*(r)r^2 dr \int R_{NL}^*(r') r' R_I(r')r'^2 dr'}{M_I-E_{NL}}\boldsymbol{\epsilon}_k\langle\pi|E^a_l \partial_l B^a_k |0\rangle,
\end{array}\label{ecc}
\end{equation}
where $\boldsymbol{\epsilon}$ is the polarization vector of $D^*$,  $R_I$, $R_F$ and $R_{NL}$ are the radial wave functions of
the initial, final and intermediate hybrid state, respectively.

The radial wave functions are calculated via solving the relativistic Schr\"odinger equation \cite{Liu:2013maa}. The potentials for the
initial and final $D^{(*)}$ mesons are taken from Ref.(\cite{Liu:2013maa}) and the potential for the intermediate hybrid states is taken from Ref.(\cite{Ke:2007ih}).

The matrix element $\langle\pi|g_E g_M E^a_l \partial_l B^a_k |0\rangle$ is of the form\cite{Kuang:1988bz,Kuang:1990kd}
\begin{equation}
\begin{array}{rl}
\langle\pi|g_E g_M E^a_l \partial_l B^a_k |0\rangle=\frac{1}{12}K_k\frac{g_E g_M }{\alpha_s}
\frac{4\pi}{\sqrt 2}\frac{m_d-m_u}{m_d+m_u}f_\pi m_\pi^2,
\end{array}
\end{equation}
where $g_E$ and $g_M$ are the coupling constants for the color electric field and color magnetic field, $\textbf{k}$ is the momentum of $\pi^0$.

In order to compare the results with the effective coupling constant $g_{D^*D\pi}$ obtained by using the QPC model, the transition amplitude of
$D^{*+} \to D^+ \pi^0$ can be rewritten as\cite{Casalbuoni:1996pg}
\begin{equation}
\begin{array}{rl}
\mathcal{M}(D^{*+} \to D^+ \pi^0)&=g^{(ME)}_{D^*D\pi} \frac{2m_D}{f_\pi}\textbf{k}\cdot\boldsymbol{\epsilon}.
\end{array}
\end{equation}
$g^{(ME)}_{D^*D\pi}$ is the effective coupling constant obtained by means of QCDME, it is then
\begin{equation}
\begin{array}{rl}
g^{(ME)}_{D^*D\pi}=\frac{1}{18 m} f_{1110}^{111}
\frac{g_E g_M}{\alpha_s}\frac{\pi}{3\sqrt2}\frac{m_u-m_d}{m_u+m_d} f_\pi m_\pi^2\sqrt{2 m_{D^*}}\sqrt{2m_D}
\frac{f_\pi}{2m_D},
\end{array}
\end{equation}
with
\begin{equation}
\begin{array}{rl}
f_{1110}^{111}=\sum_{NL}\frac{\int R_{F}(r) r R_{NL}^*(r)r^2 dr \int R_{NL}^*(r') r' R_I(r')r'^2 dr'}{M_I-E_{NL}}.
\end{array}
\end{equation}
%Then we can get the total decays by integrating out the invariant mass square $s_{K\pi}$ and $s_{\pi\pi}$.
%In next section we will present the numerical results.

%\section{Phenomenology and numerical results}

For our numerical analysis, the input parameters are taken from Ref.\cite{Agashe:2014kda},
here $m_{D}=1.869$ GeV, $m_{D^*}=2.010$ GeV, $m_\pi=0.135$ GeV, $m_c=1.800$ GeV, $m_u=0.3$ GeV,
$\frac{m_d-m_u}{m_d+m_u}=\frac{1}{3}$, $f_{B^*}=0.230$GeV, $f_K=0.160$ GeV, $F_\pi=0.093$ GeV and
$f_\pi=\sqrt 2F_\pi$. $\alpha_s=0.31$ for $\sqrt s=2.010$ GeV. Following Ref. \cite{Yan:1980uh,Kuang:1981se,Kuang:1988bz,Kuang:1990kd} we set
$\alpha_E=\frac{g_E^2}{4\pi}$, $\alpha_M=\frac{g_M^2}{4\pi}$ with $\alpha_E=0.6$ and
for a possible error range, according to the literature, we let $\alpha_M$ vary from  $\alpha_E$ to $10\alpha_E$.
The constants $f_{1110}^{111}$ and effective coupling constant $g$ obtained in terms of QCDME are listed in Tab.\ref{cc}.
%Here we take the internal hybrid states' principal quantum number $N$ to $1$ $2$ and $3$.
\begin{center}
\begin{table}[h]
\begin{tabular}[c]{|l|l|l|l|l|}
\hline
                 & $\alpha_M=\alpha_E$ &$\alpha_M=3\alpha_E$ &$\alpha_M=10\alpha_E$&$\alpha_M=30\alpha_E$ \\\hline
$f_{1110}^{111}$& 5.677 &9.833   &17.952  &31.094  \\\hline
$g_{D^*D\pi}(QCDME)$         &  0.00145    & 0.00251  &0.00459  &0.00794  \\\hline
\end{tabular}
\caption{coupling constant $g^{(ME)}_{D^*D\pi}$ and $f_{1110}^{111}$ (in units of GeV$^{-3}$), and   we set $\alpha_M$ to be $\alpha_E$, $3\alpha_E$, $10\alpha_E$,
and $30\alpha_E$ separately, and
the value of $m$ is set to $\frac{m_c}{2}$.\label{cc}}
\end{table}
\end{center}
%$g$         &  0.041    & 0.071  &0.130  &0.225  \\\hline
In order to  explore possible validity ranges of QCDME,  we extend the maximum value of $\alpha_M$ to $30\alpha_E$. One can see
that when $\alpha_M$ takes the value $30\alpha_E$,  $g^{(ME)}_{D^*D\pi}$ reaches 0.00794.
With possible errors, this result
is 60 times smaller than that obtained by the QPC model\cite{Agashe:2014kda,Meng:2008bq}.

\begin{center}
\begin{table}[h]
\begin{tabular}[c]{|l|l|l|l|l|}
\hline
                                & $\alpha_M=\alpha_E$ &$\alpha_M=3\alpha_E$ &$\alpha_M=10\alpha_E$&$\alpha_M=30\alpha_E$ \\\hline
$\Gamma(D^{*+} \to D^+ \pi^0)$ with $m=\frac{m_c}{2}$&$4.43\times10^{-5}$ &$5.31\times10^{-4}$&$4.44\times10^{-4}$ &$5.31\times10^{-3}$   \\\hline
$\Gamma(D^{*+} \to D^+ \pi^0)$ with $m=m_c     $&$1.77\times10^{-4}$&$1.33\times10^{-4}$&$1.78\times10^{-3}$&$1.33\times10^{-3}$ \\\hline
\end{tabular}
\caption{Decay width of $\Gamma(D^{*+} \to D^+ \pi^0)$(in unit of keV) from QCDME contribution, for $\alpha_M$ we takes $\alpha_E$, $3\alpha_E$, $10\alpha_E$,
 and $30\alpha_E$ respectively.
\label{dec}}
\end{table}
\end{center}
%$\Gamma(D^{*+} \to D^+ \pi^0)$ with $m=m_c$        &$4.43\times10^{-11}$ &$5.31\times10^{-10}$&$4.44\times10^{-10}$ &$5.31\times10^{-9}$   \\\hline
%$\Gamma(D^{*+} \to D^+ \pi^0)$ with $m=\frac{m_c}{2}$&$1.77\times10^{-10}$&$1.33\times10^{-10}$&$1.78\times10^{-9}$&$1.33\times10^{-9}$ \\\hline

In Tab.\ref{dec} we also list decay width $\Gamma(D^{*+} \to D^+ \pi^0)$ from QCDME contribution, from this table we can see
when $\alpha_M$ takes the maximum value $30\alpha_E$, the decay width of $\Gamma(D^{*+} \to D^+ \pi^0)$ from QCDME contribution
is $5.31\times10^{-3}$keV. This result
is more than three orders smaller than the result obtained by using the QPC model($21.9$keV).

\section{conclusion and discussion  }

% Eichten et al's statement that above the production threshold of states which have different quantum numbers
%(except the principal numbers),
Our numerical results for decay mode $D^*\to D\pi^0$ confirm that the direct pion emission is not the leading term and the contributions determined by the QCDME must be much
smaller than that induced by other mechanisms.

Let us try to understand the smallness of the contribution from direct pion emission, namely using data to convince that our estimate is
reasonable.
There are two suppression factors: the one-pion emission violates the isospin conservation and
QCDME is an OZI suppressed mechanism. Comparing with the vacuum quark pair creation, it must be small and we see that its contribution to the decay width ranges
about $10^{-4}\sim 10^{-5}$ keV. As we know, the decay $\Psi(2S)\to J/\Psi +\pi^0$
is also an  isospin violation and OZI suppressed process, its branching ratio is $1.27\times 10^{-3}$\cite{Agashe:2014kda}, slightly larger than
our estimate for $D^*\to D\pi^0$. That further suppression factor is coming from the fact that the
transition $\Psi(2S)\to J/\Psi +\pi^0$ is an $S-$wave process, whereas $D^{*+} \to D^+ \pi^0$ is a $P-$wave one whose decay width
is proportional to the three-momentum ${\bf k}$ which is
small, so the decay width $\Gamma(D^{*+} \to D^+ \pi^0)$ suffers from a P-wave suppression.

We may also look at $\psi(2S)$ as an example for a direct $\pi$ emission. $\psi(2S)\to \pi^0+h_c(1P)$ can only occur via a direct pion emission, so is
completely determined by the QCDME mechanism, and its partial width is about 0.26 keV. Since it is an S-wave process, it does not suffer from the P-wave suppression.
In $\psi(2S)$ decays, the mode $\psi(2S)\to \eta_c+\pi^0$ is not seen, but $\psi(2S)\to\eta_c+\pi^+\pi^-\pi^0$ has been measured, and its branching ratio is not too
small ($< 1.0\times 10^{-3}$), that is because the direct emission of three pions does not violate the isospin conservation.

Equivalently, the QCDME can be replaced by the chiral perturbation
theory, for example the transition of $\Upsilon(nS)\to
\Upsilon(mS)+\pi^+\pi^-$ was studied in terms of the chiral
theory\cite{Guo:2006ai}.

Moreover, we have extended our mechanism to study the non-resonant three-body decays of $B$ where the weak interaction gets involved. The contribution
has been studied by Cheng et al. in terms of the chiral perturbation theory and we have re-done the evaluation by means of QCDME.
We will present the relevant results in our next work.

As a conclusion, we confirm the validity of the QCDEM and determine its application region. It is indicated that
since the framework applies only to the direct pion emission, if there are other mechanisms to
contribute, such as the quark-pair creation from vacuum to  $D^*\to D+\pi^0$; or $\Upsilon(5S)\to B\bar B^*\to \Upsilon(1S)+\pi^+\pi^-$ where an intermediate
state of $B\bar B^*$ portal is open, i.e the available energy is above the production threshold of $B\bar B^*$, the direct emission induced by the QCDME is no longer
the leading term and can only contribute tiny fraction.

\section*{Acknowledgments}

We greatly benefit from the talk given by Prof. Eichten at Nankai
University and we also sincerely thank Prof. H.Y. Cheng for
enlightening discussions. This  work is supported by National
Natural Science Foundation of China under the Grant Number
11375128.

\appendix*

\section{Some formulae}
In the $^3P_0$ model, for a two-body decay process $A \to BC$, the total wave function of a meson  can be  written as
{\small
\begin{eqnarray}\label{mockmeson}
&&\left|A(n_A \mbox{}^{2S_A+1}L_A \,\mbox{}_{J_A M_{J_A}})
({\textbf{K}}_A) \right\rangle \nonumber\\&&= \sqrt{2
E_A}\sum_{M_{L_A},M_{S_A}} \left\langle L_A M_{L_A} S_A M_{S_A} |
J_A M_{J_A} \right\rangle \nonumber\\&&\quad\times\int \rm d
\mathbf{k}_1\rm
d\mathbf{k}_2\delta^3\left(\textbf{K}_A-\mathbf{k}_1-{\mathbf{k}}_2\right)\Psi_{n_A
L_A M_{L_A}}\left(\mathbf{k}_1,\mathbf{k}_2\right) \nonumber\\
&&\quad\times\chi^{1 2}_{S_A M_{S_A}}\varphi^{1 2}_A\omega^{1 2}_A
\left|\;q_1\left(\mathbf{k}_1\right)
\bar{q}_2\left(\mathbf{k}_2\right)\right\rangle,
\end{eqnarray}}
which satisfies the normalization condition: $\langle A(\textbf{K}_A)|A(\textbf{K}'_A) \rangle= 2E_A\,\,\delta^3(\textbf{K}_A-\textbf{K}'_A).$ The subscripts 1 and 2 refer to the quark and anti-quark within meson $A$. ${\textbf{K}}_A$ is the momentum of the meson $A$. $\chi^{1 2}_{S_A M_{S_A}}, \varphi^{1 2}_A, \omega^{1 2}_A$ are the spin, flavor and color parts respectively. $\Psi_{n_A L_A M_{L_A}}\left(\mathbf{k}_1,\mathbf{k}_2\right)$ is the  spatial part of a meson wavefunction in the momentum representation. For the concerned mesons are in the ground states, the simple harmonic oscillator (HO) wavefunction is employed which reads as
\begin{eqnarray}
\Psi_{00}(\mathbf{k})&=&\frac{1}{\pi^{3/4}}R^{3/2}\exp\left(-\frac{R^2\mathbf{k}^2}{2}\right),
%\Psi_{1\mu}(\mathbf{k})&=&i\frac{\sqrt{2}}{\pi^{3/4}}R^{5/2}k_\mu\exp\left(-\frac{R^2\mathbf{k}^2}{2}\right),
\end{eqnarray}
where $\mathbf{k}=(m_2\mathbf{k}_1-m_1\mathbf{k}_2)/(m_1+m_2)$ is the relative momentum between the
quark 1 (with mass $m_1$) and the anti-quark 2 (with mass $m_2$)  within a meson. The $R$ value is fixed by fitting experimental data.

The explicit matrix element is depicted as

\begin{eqnarray}\label{T-matrix}
&&\langle BC|T|A\rangle=\sqrt{8 E_A E_B
E_C}\;\;\gamma\!\!\!\!\!\!\!\!\!\!\!
\sum_{\renewcommand{\arraystretch}{.5}\begin{array}[t]{l}
\scriptstyle M_{L_A},M_{S_A},\\
\scriptstyle M_{L_B},M_{S_B},\\
\scriptstyle M_{L_C},M_{S_C},m
\end{array}}\renewcommand{\arraystretch}{1}\!\!\!\!\!\!\!\!
\langle 1\;m;1\;-m|\;0\;0 \rangle\nonumber\\&&\quad\times \langle
L_A M_{L_A} S_A M_{S_A} | J_A M_{J_A} \rangle\langle L_B M_{L_B}
S_B M_{S_B} | J_B M_{J_B} \rangle\nonumber\\&&\quad\times\langle
L_C M_{L_C} S_C M_{S_C} | J_C M_{J_C} \rangle\langle\varphi^{1
3}_B \varphi^{2 4}_C | \varphi^{1 2}_A \varphi^{3 4}_0
 \rangle\nonumber\\&&\quad\times
 \langle \chi^{1 3}_{S_B M_{S_B}}\chi^{2 4}_{S_C
M_{S_C}}  | \chi^{1 2}_{S_A M_{S_A}} \chi^{3 4}_{1 -\!m} \rangle
I^{M_{L_A},m}_{M_{L_B},M_{L_C}}({\textbf{K}}) \;,\nonumber
\end{eqnarray}
where  $I^{M_{L_A},m}_{M_{L_B},M_{L_C}}(\textbf{K})$ is a spatial integral,
reading as
\begin{eqnarray}
&&I^{M_{L_A},m}_{M_{L_B},M_{L_C}}(\textbf{K}) = \int\!\rm
d\mathbf{k}_1\rm d\mathbf{k}_2\rm d\mathbf{k}_3\rm
d\mathbf{k}_4\,\delta^3(\mathbf{k}_1+\mathbf{k}_2)\nonumber\\
&&\quad\times\delta^3(\mathbf{k}_3+\mathbf{k}_4)\delta^2
(\textbf{K}_B-\mathbf{k}_1-\mathbf{k}_3)\delta^3(\textbf{K}_C-\mathbf{k}_2-\mathbf{k}_4)\nonumber\\
&&\quad\times\Psi^*_{n_B L_B
M_{L_B}}(\mathbf{k}_1,\mathbf{k}_3)\Psi^*_{n_C L_C
M_{L_C}}(\mathbf{k}_2,\mathbf{k}_4)\nonumber\\&&\quad\times
\Psi_{n_A L_A M_{L_A}}(\mathbf{k}_1,\mathbf{k}_2)
\mathcal{Y}_{1m}\Big(\frac{\mathbf{k}_3-\mathbf{k}_4}{2}\Big).
\label{integral}
\end{eqnarray}

\end{document}